\documentclass{aa}  
\bibpunct{(}{)}{;}{a}{}{,} % to follow the A&A style 

\usepackage{graphicx}
\usepackage{txfonts}
\usepackage{natbib,twoopt}
\usepackage[draft]{hyperref} %% to avoid \citeads line fills
\bibpunct{(}{)}{;}{a}{}{,}             %% natbib format for A&A and ApJ
\makeatletter
  \newcommandtwoopt{\citeads}[3][][]{\href{http://adsabs.harvard.edu/abs/#3}%
    {\def\hyper@linkstart##1##2{}%
     \let\hyper@linkend\@empty\citealp[#1][#2]{#3}}}
  \newcommandtwoopt{\citepads}[3][][]{\href{http://adsabs.harvard.edu/abs/#3}%
    {\def\hyper@linkstart##1##2{}%
     \let\hyper@linkend\@empty\citep[#1][#2]{#3}}}
  \newcommandtwoopt{\citetads}[3][][]{\href{http://adsabs.harvard.edu/abs/#3}%
    {\def\hyper@linkstart##1##2{}%
     \let\hyper@linkend\@empty\citet[#1][#2]{#3}}}
  \newcommandtwoopt{\citeyearads}[3][][]%
    {\href{http://adsabs.harvard.edu/abs/#3}
    {\def\hyper@linkstart##1##2{}%
     \let\hyper@linkend\@empty\citeyear[#1][#2]{#3}}}
\makeatother

\begin{document}

   \title{TEE, a simple estimator for the precision of eclipse and transit minimum times}

   \author{H. J. Deeg
          \inst{1,2}
          \and
          B. Tingley\inst{3}
          }

   \institute{Instituto de Astrof\'\i sica de Canarias, C. Via Lactea S/N, E-38205 La Laguna, Tenerife, Spain
   \and
Universidad de La Laguna, Dept. de Astrof\'\i sica, E-38206 La Laguna, Tenerife, Spain\\
              \email{hdeeg@iac.es}
         \and
        Stellar Astrophysics Centre (SAC), Department of Physics and Astronomy, Ny Munkegade 120, 8000 Aarhus C, Denmark\\
             \email{tingley@phys.au.dk}
             }

   \date{Received XXXX; accepted XXXX}

% \abstract{}{}{}{}{} 
% 5 {} token are mandatory
 
  \abstract
  % context heading (optional) %leave it empty if necessary  
  {Transit or eclipse timing variations have proven to be a valuable tool in exoplanet research. However, no simple way to estimate the potential precision of such timing measures has been presented yet, nor are guidelines available regarding the relation between timing errors and sampling rate.} 
   { A timing error estimator (TEE) equation is presented that requires only basic transit parameters as input. With the TEE,  estimating timing precision  for actual data and for future instruments, such as the TESS and PLATO space missions, is straightforward.}
   {A  derivation of the timing error based on a trapezoidal transit shape is given. We also verify the TEE on realistically modelled transits using Monte Carlo simulations and determine its validity range, exploring in particular the interplay between ingress/egress times and sampling rates.}
   {The simulations show that the TEE gives timing errors very close to the correct value, as long as the temporal sampling is faster than transit ingress/egress durations and transits with very low S/N are avoided.}
   {The TEE is a useful tool for estimating eclipse or transit timing errors in actual and future data sets. In combination with a previously published equation to estimate period-errors, predictions for the ephemeris precision of long-coverage observations are possible as well. The tests for the TEE's validity range also led  to implications for instrumental design. Temporal sampling has to be faster than transit ingress or egress durations, or a loss in timing precision will occur. An application to the TESS mission shows that transits close to its detection limit will have timing uncertainties that exceed 1 hour within a few months of their acquisition. Prompt follow-up observations will be needed to avoid 'losing' their ephemerides.}


   \maketitle
%
%____________________________________________________________
\section{Motivation for an estimator of eclipse and transit timing errors}
The precise timing of stellar eclipses and planetary transits has a long track record in astronomy, and numerous methods have been developed to measure the times of minimum brightness in eclipse-like events \citepads[e.g.]{1956BAN....12..327K,1967AJ.....72..226W,2006Ap&SS.304..363M,2012A&A...540A..62O}. Eclipse minimum times, translated into variations of orbital periods, have been widely used in the analysis of eclipsing binary systems. More recently they  also became important in the analysis of light curves from transiting exoplanets. In this case, deviations from strict periodicity, known as transit timing variations (TTVs), have usually been taken as  indications of the presence of additional bodies in the system \citepads[e.g.][]{2005MNRAS.359..567A, 2005Sci...307.1288H, 2007MNRAS.377.1511H}. In such cases, it is important to ensure that a supposed TTV\footnote{In the remainder of this work, we use the terms `transit' and `eclipse' interchangeably.} is really due to an astrophysical effect and not an artefact of observational noise. This requires a technique to estimate  the precision of the measured transit ephemerides. This need for an estimation of timing and ephemeris errors also arose during the ground-based photometric follow-up of planet candidates from the CoRoT space mission \citepads{2009A&A...506..343D}%deeg09
. In this case, a reliable estimator of the timing error was necessary to determine when and if follow-up photometric observations of a transit candidates should be scheduled:  if the expected timing error exceeded a few hours, the observations would not be conclusive. During the course of the mission, we became aware that the errors in the ephemerides given to us,  which had been derived by other members of the CoRoT-team,  varied by factors of  `a few' for candidates with similar basic parameters, like brightness, transit depth, and transit duration. Their ephemerides and their errors had been derived via a variety of methods, mostly by the fitting of different types of models -- from trapezoid-shaped to more elaborate ones -- but also eye estimates were used on occasion.
A simple and reliable estimator for deriving timing errors was therefore desirable in order to perform follow-up observations with little delay, which led to the method presented here. In combination with the estimator for period-errors presented in \citetads{2015A&A...578A..17D}, predictions for the time-uncertainty of future transits,   which are essential for any follow-up observations,  could now be performed quickly.

%comparisons to other work
This work is distinct from other, similar works in the literature. The original work on the subject, which in part inspired this work, appeared in \citetads{2004IAUS..213...80D}. In that work the authors describe the timing error for triangular-shaped binary eclipses. This was later extended by \citetads{2010MNRAS.405..657S}%sybilski2010
, who explored the detectability of TTVs in binary star systems for the purpose of detecting circumbinary planets. They used a single eclipsing binary model and a single set of observing parameters (integration time, noise) for CoRoT and Kepler simulations and a few for ground-based simulations, ultimately verifying the equations presented in \citetads{2004IAUS..213...80D}. \citetads{2008ApJ...689..499C} present analytic approximations for transit light-curve observables, among them an equivalent equation for the timing uncertainty. \citetads{2013AJ....145..148M} and \citetads{2015A&A...578A..17D} discuss ephemeris errors from continuous sets of timing measurements covering multiple orbital periods, as given by data from space missions like CoRoT and Kepler. These  two papers, however, do not contemplate the expected timing precision of  {individual} transit events, whose validation is the core of the present work. 

Combining the work presented here and that by \citetads{2015A&A...578A..17D} provides  the tools for  deriving both the epoch and the period-error of any linear ephemerides that are based on continous-coverage data, needing only basic eclipse or transit parameters. We expect this work to be of particular interest for the study of the significance of transit timing variations (TTVs) and for the evaluation of the timing precision that can be expected in future instruments, such as the upcoming TESS and PLATO missions.

%Our work
In the following, we present the formulae to derive timing errors of individual transits, which we refer to as the timing error estimator (TEE). By using a trapezoidal approximation for the transit shape, it is possible to derive an analytical expression for the timing error of an individual transit based on a few basic parameters. We concentrate on the derivation of the timing error of individual eclipses and extend this by assessing the impact of sampling rate on the timing error. We begin by exploring trapezoidal model eclipses to verify that the expression works as expected, then extend it to real modelled exoplanetary transits.

\section{Eclipse timing error estimator, TEE}

 \begin{figure}
   \centering
   \includegraphics[width=9.5cm]{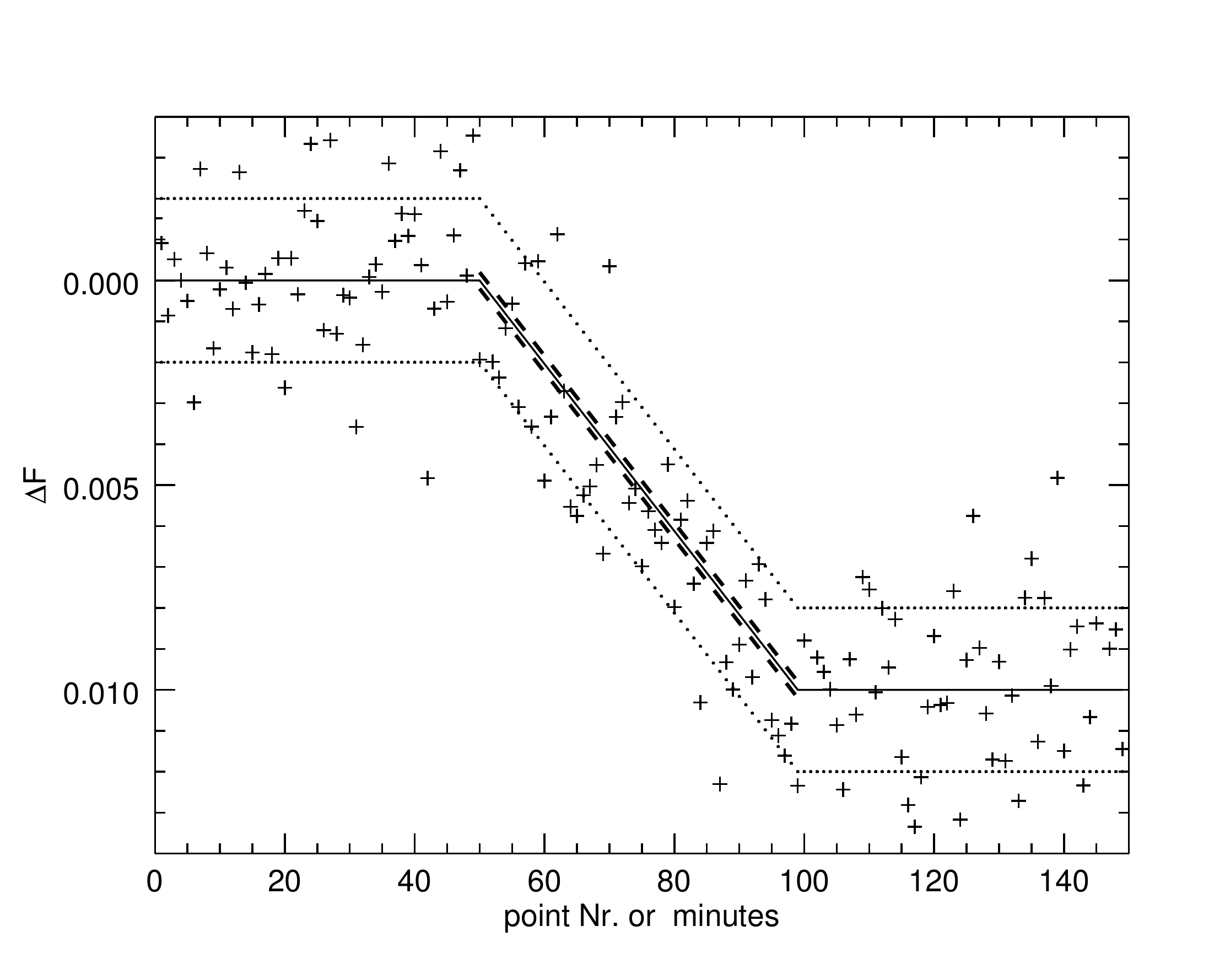}
  
   \caption{Model of an eclipse ingress of $\Delta F =0.01$ (solid line) with simulated data (crosses). The dotted lines outline the $\sigma_{F_\mathrm{rms}}=0.002$ standard deviation of the individual data points, and the dashed lines outline the expected error of the average over all 50 points within the ingress.}
     \label{fig:ingress}
    \end{figure}

The derivation of TEE arises from the use of a trapezoid to approximate the shape of the eclipse where the flat part is of any length, including zero.
Furthermore, it assumes that the parameters of the trapezoid (amplitude and durations of ingress, flat part and egress) are known with negligible errors, such that the only free parameter is the eclipse mid-time. Given these assumptions, let us consider the ingress of an eclipse (Fig.~\ref{fig:ingress}), with a depth of 1\% or $\Delta F =0.01$. The observations are simulated by taking a model trapezoid and adding white noise, at a level of $\sigma_{F_\mathrm{rms}}=0.002$ in the given case. In real observations, the level of white noise would be derived from out-of-eclipse data. The white noise level during the ingress is then $\sigma_{F_\mathrm{in}} = \sigma_{F_\mathrm{rms}}/\sqrt{N_\mathrm{in}}$, where $N_\mathrm{in}$ is the number of points during ingress. In the event that the noise does not scale in this fashion, e.g. due to strong red noise, then $\sigma_{F_\mathrm{in}}$ should be determined by other means. The levels of $\pm \sigma_{F_\mathrm{in}}$ relative to the trapezoid are shown as dashed lines in Fig.~\ref{fig:ingress}. These lines also correspond to the expected 1$\sigma$ errors when fitting the known model ingress to the data when the only free parameter is the time of ingress. The slope $T_\mathrm{in}/{\Delta F}$ then determines  the corresponding error in time, given as $\sigma_{t_\mathrm{in}}= \sigma_{F_\mathrm{in}}{T_\mathrm{in}}/{\Delta F}$.

Considering that observations will generally contain both ingress {and} egress, the only change is that instead of using only $\sigma_{F_\mathrm{Tin}}$, we should use the photometric noise on the  timescale of ingress and egress summed, $\sigma_{F_\nabla}$, with $T_\nabla$ being the summed duration of ingress and egress (the $\nabla$ symbol denotes ingress plus egress, and is motivated by their triangular shape). This leads to the most basic form of the TEE\footnote{While Eq.~\ref{eq:TEEbasic} uses relative values for the photometric noise $\sigma_{F}$ and the transit amplitude $\Delta F$, we note that for small  amplitudes $\Delta F$, the use of magnitude units for both parameters will lead to identical results}

\begin{equation}
\sigma_\mathrm{t}=  \frac{\sigma_{F_\nabla}\ T_\nabla}{ 2\,\Delta F}  
\label{eq:TEEbasic}
,\end{equation}
where $\sigma_\mathrm{t}$ denotes the timing precision and $T_\nabla \approx 2 T_\mathrm{in}$ was used.  

We  also define a timing signal-to-noise, $(S/N)_t$, given by 
\begin{equation}
(S/N)_t = 2\Delta F/\sigma_{F_\nabla}.
\label{eq:SNt}
\end{equation}
The relative timing error, $\sigma_\mathrm{t}/T_\nabla$, is then  simply the inverse of $(S/N)_t$:

\begin{equation}
\frac{\sigma_\mathrm{t}}{T_\nabla}=  \frac{1}{(S/N)_t}  
\label{eq:TEEnorm}
.\end{equation}

Normally, the noise of the data on the timescale $T_\nabla$ is not well known, but will be known on another timescale $\tau$ (e.g. over the sampling time between  data points). We propose, therefore, a white noise scaling from the known noise $\sigma_{F_\tau}$ over the timescale $\tau$ to the noise $\sigma_\mathrm{F_\nabla}$ using

\begin{equation}
\sigma_\mathrm{F_\nabla} = \sigma_{F_\tau} \ \sqrt{\tau/T_\nabla}
\label{eq:noisescale}
,\end{equation}
where we assume that the number of data points is proportional to the length of the timescales $\tau$ and $T_\nabla$.
This leads then to this general form of the TEE:
\begin{equation}
\sigma_\mathrm{t}= \frac{\sigma_{F_\tau}}{ 2\, \Delta F} \sqrt{\tau \ T_\nabla}
\label{eq:TEEgen}
.\end{equation}
For convenience, we also provide this equation in terms of the ingress time $T_\mathrm{in} = T_\nabla /2$: 
\begin{equation}
\sigma_\mathrm{t}= \frac{\sigma_{F_\tau}}{\Delta F} \sqrt{\frac{\tau \ T_\mathrm{in}}{2}}
\label{eq:TEEtin}
.\end{equation}

In real time-series data, conversions between noises on different timescales often deviate strongly from white noise scalings when timescales differing by more than an order of magnitude are considered. It is therefore advisable to characterise the noise on a timescale that is not very different from $T_\nabla$ when using TEE. For example, when we predicted the timing precision of CoRoT transits, we used the photometric noises over a timescale of $\tau$=2 hours given by  \citetads{2009A&A...506..425A}%Aigrain
. This timescale is suitable as it is of the same order of magnitude as the $T_\nabla$ of the transit candidates. On the other hand, a scaling from the point-to-point errors of CoRoT data would lead to an underestimation of the timing precision, due to a significant red noise component on a two-hour timescale relative to CoRoT's standard sampling time of 512 seconds \citepads{2009A&A...506..425A}.
However, if a white noise scaling from the point-to-point noise, $\sigma_{F_\mathrm{rms}}$, to $\sigma_{F_\nabla}$ is acceptable, we may also use a version of the TEE in terms of the number of points during ingress-plus-egress, $n_\nabla$, with $n_\nabla=T_\nabla / \tau$ (here $\tau$ is the time between exposures):

\begin{equation}
\sigma_\mathrm{t}=\frac{\sigma_{F_\mathrm{rms}}\  T_\nabla}{2\ \Delta F \sqrt{n_\nabla}}
\label{eq:tee_rms}
.\end{equation}
 
We note that the above equation is well-suited for Kepler data since no relevant red noise \citepads{2010ApJ...713L.120J} affects the scaling from the nominal 30-minute sampling time of Kepler long-cadence data to the hour-long ingress/egress durations of most planet transits, represented by $T_\nabla$ .   
 
 \begin{figure}
   \centering
   \includegraphics[width=9.5cm]{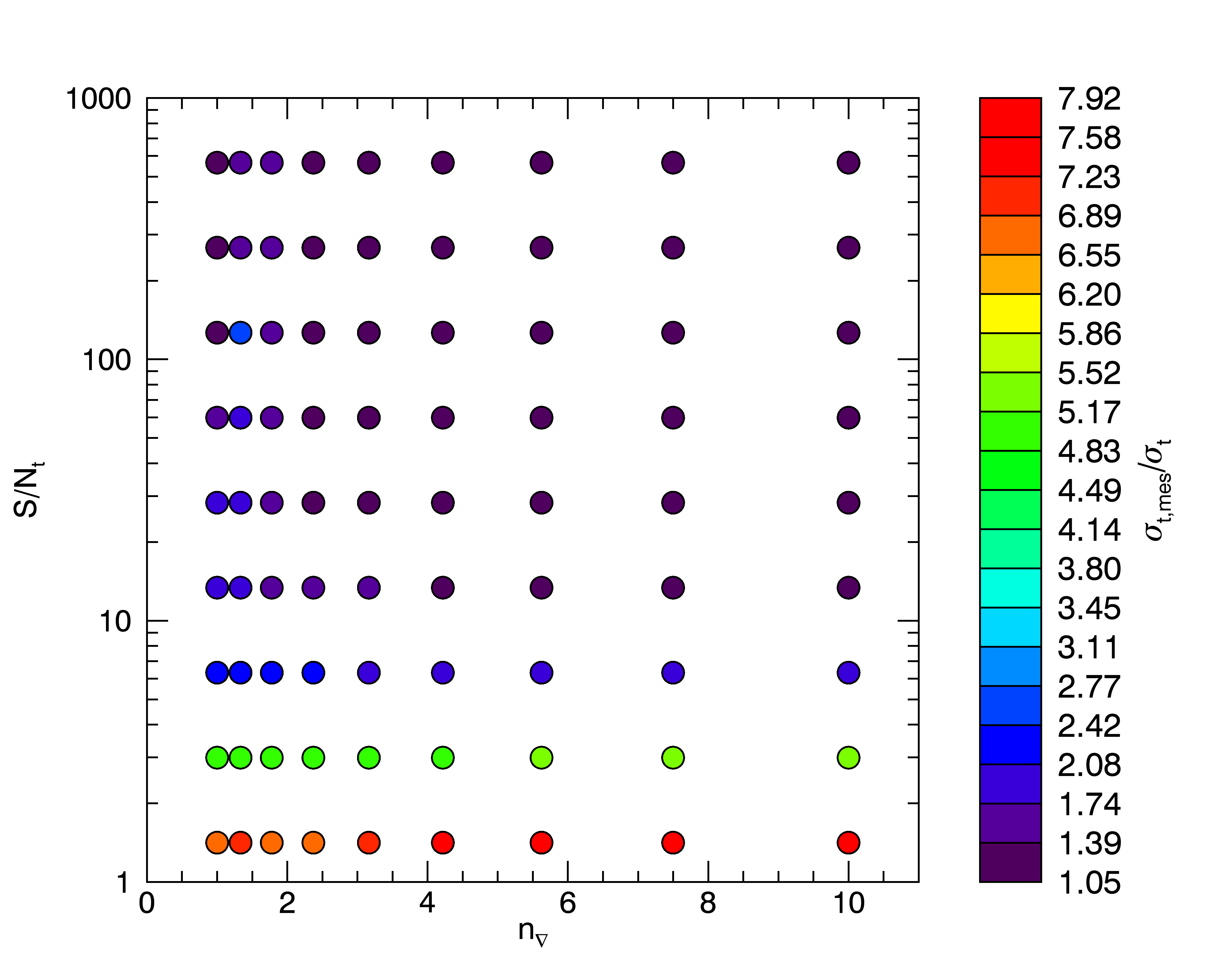}
  
   \caption{Performance of the TEE based on simulations of trapezoid-shaped transits. The colour of each point indicates the ratio between the measured timing precision and the precision predicted by the TEE, $\sigma_\mathrm{t, meas}/\sigma_\mathrm{t}$, on the scale shown to the right. The measured precision has been derived from 10000 injection and retrieval simulations; see the text for details. The axes correspond to the simulations' input variables, which are the signal-to-noise relevant for timing precision (see Eq.~\ref{eq:SNt}) and $n_\nabla$, which is the combined number of points during ingress and egress.}
              \label{fig:mes_vs_teo_trapez}
    \end{figure}
    
\section{Scope and limits of the TEE}
This theoretical equation requires a round of practical tests in order to demonstrate its validity. To this end, we perform two sets of Monte Carlo simulations based on model injection and parameter retrieval. First, we tested sets of modelled trapezoidal transits like those used to derive the TEE. Then we did the same for sets of model transits like those we expect to encounter in nature. In both cases, we performed Monte Carlo simulations of these models to assess how well we could recover the transit timing. For this we fitted the known eclipse models to find eclipse mid-times and then subtracted these from the input mid-times, resulting in timing errors for each individual simulation. For each set of parameters, `measured' timing errors $\sigma_\mathrm{t, meas}$ were then derived from the variance of timing errors from 10000 simulations. We then compared the measured errors to $\sigma_\mathrm{t}$, the timing errors expected from the TEE, via the ratio $\sigma_\mathrm{t, meas}/\sigma_\mathrm{t}$.

\subsection{Behaviour of the TEE for trapezoid-shaped transits}
The use of a trapezoidal shape in the first sets of simulations was motivated by the use of that same shape for the derivation the TEE in order to determine the TEE's fundamental correctness and range of validity. The predicted relative timing error $\sigma_\mathrm{t}/T_\nabla$ depends only on one parameter, $(S/N)_t$ (see Eq.~\ref{eq:SNt}), but the range of $(S/N)_t$ for which the TEE is reliable was uncertain. Also uncertain was the range of validity of the TEE against another parameter that can be expected to affect the timing precision, the number of data points during ingress and egress, $n_\nabla$. 

We used a Monte Carlo approach to obtain the $\sigma_\mathrm{t, meas}$. For each of the 10000 Monte Carlo trials, we generated time series with a specified sampling rate and level of white noise, and then inserted a trapezoidal event with a given value of $n_\nabla$ and $(S/N)_t$ at a random time $t_{\rm c}$ safely away from either end of the time series. Then, assuming {a priori} knowledge of the shape of the trapezoid, we determine its mid-time $t_{\rm c}$ from a simple $\chi^2$ minimisation. We obtained a $\Delta t_{\rm c}$ from each trial by taking the difference between the input and output $t_{\rm c}$, keeping the $\chi^2$ as an indication of the quality of the fit. The measured timing precision $\sigma_\mathrm{t, meas}$ was then obtained from the standard deviation of the $\Delta t_{\rm c}$ from the 10000 trials. This process was then repeated for each set of input parameters $n_\nabla$ and $(S/N)_t$.

The results from these simulations are shown in Fig.~\ref{fig:mes_vs_teo_trapez}, expressed as the ratio between the measured timing precision and that predicted by the TEE, $\sigma_\mathrm{t, meas}/\sigma_\mathrm{t}$. This ratio approaches 1, meaning that the TEE performs a valid prediction of the timing precision, if two conditions are met: $n_\nabla > 2$ and $S/N_\mathrm{t} \gtrsim 10$. On the other hand, if the sampling rate or the S/N is below these limits, the real timing precision may be much worse than that given by the TEE.
 
Equation~\ref{eq:TEEnorm} shows that for  $S/N_\mathrm{t} \la  1$, no useful timing error can be measured as $\sigma_\mathrm{t}$ will exceed the length of an ingress or egress. In practice -- and shown by the simulations -- more stringent limits of $S/N_\mathrm{t}$ apply since with decreasing S/N, the reliability of eclipse detections decreases and contributions from random detections with arbitrary $\Delta t_{\rm c}$ values become relevant.

\begin{figure}
 \centering
   \includegraphics[width=9.5cm]{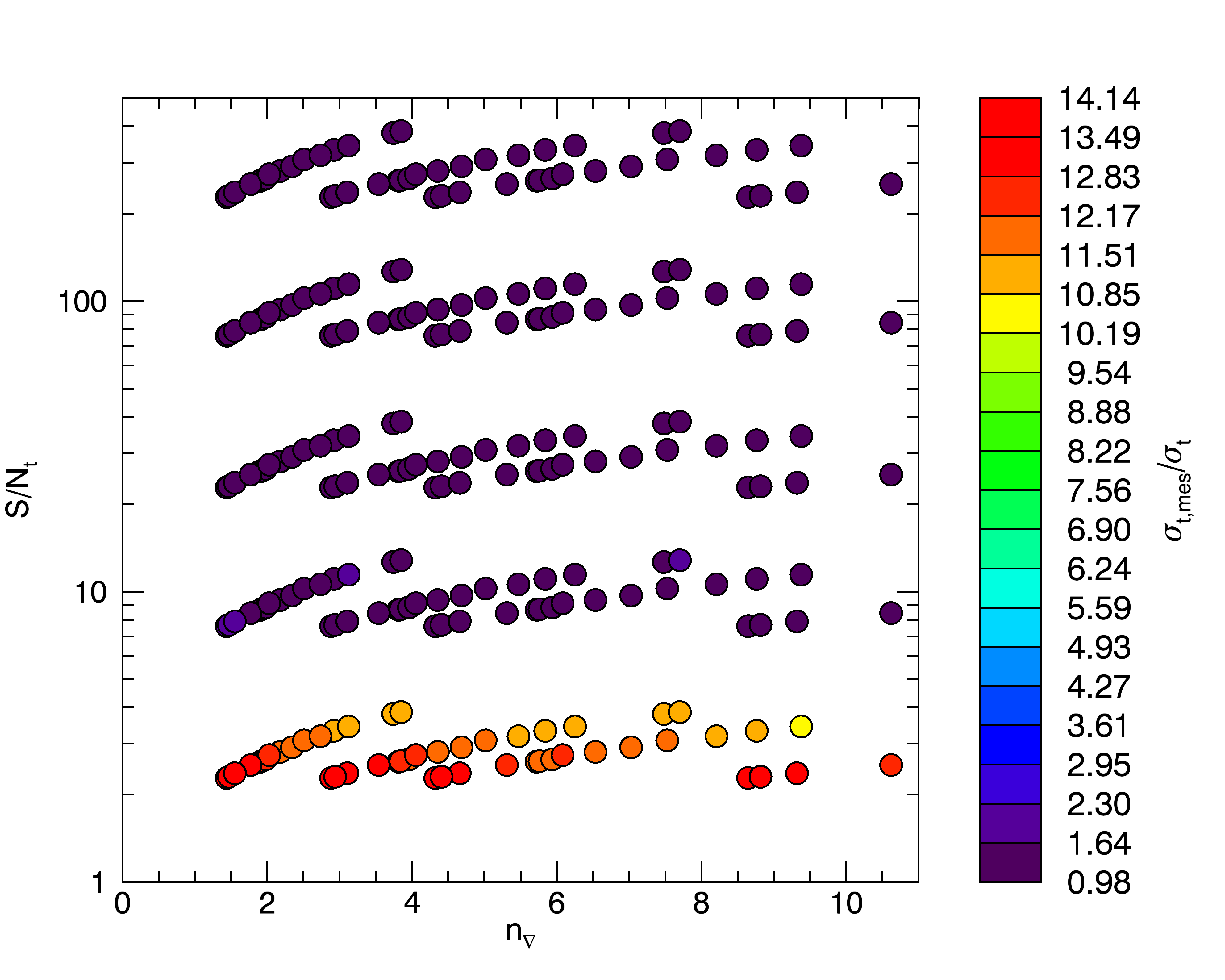}
  
   \caption{As in Fig.~\ref{fig:mes_vs_teo_trapez}, but from simulations using naturally shaped transits. The distribution of points does not adhere here to the grid apparent in Fig.~\ref{fig:mes_vs_teo_trapez} since  the two parameters ($(S/N)_t, n_\nabla$) spanning this figure were calculated a posteriori from several input parameters that were varied between simulations (limb-darkening, impact parameter, sampling rate, white noise level); see text.}
              \label{fig:mes_vs_teo_real}
    \end{figure}

\subsection{Behaviour of the TEE for naturally occurring transits}
 
Now that the TEE against trapezoid-shaped transits has been validated, we extend here the analysis to natural eclipse shapes as these are of more practical interest. These tests are more complex than for the trapezoidal models that we used previously, as values such as $T_\nabla$ and $\Delta F$ are no longer input variables for the tests and must instead be approximated in some way in order to evaluate the accuracy of TEE. In theory, it should be possible to perform a numerical analysis of a model transit in order to calculate these values exactly. However, for the purposes of this study (which only attempts to evaluate the errors in timing, but not errors in other transit parameters) it is sufficient to fit trapezoids to naturally shaped transits, as this will deliver the parameters that are relevant for the TEE. To that end, we proceeded with the following steps:\\

{\it i}) A set of noiseless model light curves was generated from pixelated representations of a stellar disc that is occulted by a transiting planet, employing the same method as the UTM transit modeller \citepads{2014ascl.soft12003D}. All light curves were generated for transits of a planet of 0.09 stellar radii across a solar-like star, while the following parameters were varied step-wise, leading to a total of 480 combinations: The impact parameter of the transit was incremented in eight steps, from an equatorial transit with b=0 to a nearly polar (but still total) transit with b=0.9. Models were generated with six different sampling rates,  where the coarsest models represented the entire transit in six points and the finest ones represented it in 360 points. All models were generated for two visual passbands, namely $V$ and $I$, using the corresponding quadratic solar limb-darkening coefficients. All of these models were then normalised such that the deepest part of the transit corresponds to a decrease in flux of $\Delta F = 1\%$. 
A further parameter, photometric white noise on five different levels, was added to the models in step {\it iii}, but is already mentioned here. This noise had values ranging from $\sigma_{F_\tau} = 0.0001$ to $0.01$, where the timescale $\tau$ corresponds to  1/6 of the total transit duration\footnote{The indicated noises correspond therefore to the point-to-point noise in the models with the coarsest sampling. For models with finer sampling, the noise was incremented correspondingly, using a white noise scaling.}.
Each point in Fig.~\ref{fig:mes_vs_teo_real} corresponds to one of the possible parameter combinations\footnote{Only about half of the 480 possible parameter combinations are shown in Fig.~\ref{fig:mes_vs_teo_real}, with the other half leading to values of $n_\nabla$ outside of the displayed range, up to $n_\nabla \approx 230$.}. \\

{\it ii}) For each of these noiseless transit models, we obtained a trapezoidal fit and converted its output into the parameters relevant for the TEE, namely $\Delta F$ and $T_\nabla$ respectivly $n_\nabla$. Since the models were those of real transits, the $\Delta F$ of the best-fit trapezoids was always lower than the 1\% flux decrease of the input models. Also, owing to the nature of the trapezoidal fit, the resulting fit duration was always longer than the transit duration of the input models.\\

{\it iii}) For each set of input parameters we performed 10\ 000 injection and retrieval trials. For each trial light curve, random white noise at one of the levels specified in step {\it i} was added to the noiseless models and the transit mid-time was offset by a random amount, which was recorded. \\

{\it iv}) On each of the trial light curves, we used the noiseless natural transit shape (obtained in step {\it i}) to perform a one-parameter fit for the transit mid-time, and recorded the difference between the fitted mid-time and the input mid-time.\\

{\it v}) The subsequent analysis proceeded similarly to that in Sect.~{3.1}. From the distribution of input-output differences from each of the 10\ 000 trials, we obtained the measured timing precision $\sigma_\mathrm{t, meas}$ and compared it to the TEE estimate $\sigma_\mathrm{t}$, which was based on the parameters from step {\it ii} and the known noise $\sigma_{F_\tau}$. This was done for each parameter set, with the ratio $\sigma_\mathrm{t, meas} / \sigma_\mathrm{t}$ determining the colour of the points in Fig.~\ref{fig:mes_vs_teo_real}.\\

We note that the use of a noiseless model in  step {\it ii}   to obtain the transit parameters except mid-time corresponds to the usual procedure employed in the analysis of long-coverage light curves. There, transit parameters are derived from phase-wrapped superpositions of multiple individual transits, resulting in a light curve with a higher S/N than that of the individual transits from which the timing measurements are  obtained.

The bulk of the sets in Fig.~\ref{fig:mes_vs_teo_real} show a ratio $\sigma_\mathrm{t, meas}/\sigma_\mathrm{t}$ close to one, demonstrating that a trapezoidal fit to the transit is sufficient to estimate the expected timing error. However, a close comparison with the results from the trapezoidal tests depicted in Fig.~\ref{fig:mes_vs_teo_trapez} shows some subtle differences: reduced detectability for low values of $(S/N)_t \la 3$, improved performance around $(S/N)_t \sim 7$, and in general a better performance for low values of $n_\nabla$. This is due to the fact that real transits have some curvature all the way through the transit. The $T_\nabla$ derived from the trapezoidal fit therefore underestimates the true effective value of this parameter, which mitigates the detrimental effects of low sampling rates. The differences for low $(S/N)_t$ may be real or may be an artefact of the fitting process. In either case,  the lowest S/N cases have timing errors 
$\sigma_\mathrm{t} $ that approach or exceed the transit duration and neither are the transits themselves reliably detected.

\section{An example application: Timing precision of TESS detections} 

\begin{table*}
\label{tab:TESStiming}
\caption{Transit timing precision and ephemeris precision of example transit detections by the TESS mission; see text for details.}
 %\centering
   \includegraphics{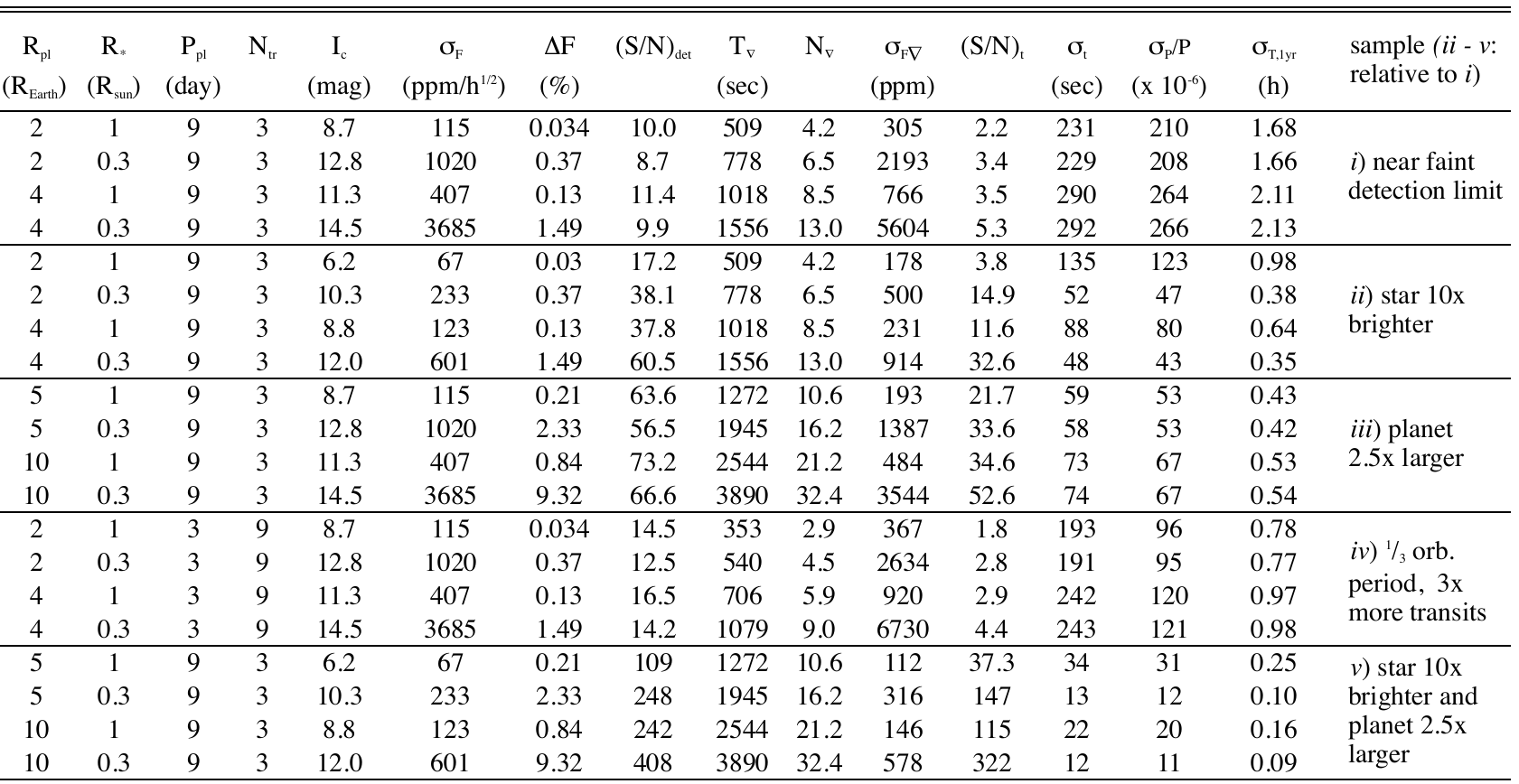}
\end{table*}
 
In the following, we show an application of the TEE towards the estimation of the timing precision of transiting planets (or planet candidates) detected by the TESS mission. TESS \citepads{2014SPIE.9143E..20R,2015JATIS...1a4003R} is a NASA mission slated for launch in late 2017 that will survey nearly the entire sky for transiting planets, with the majority of the sky being covered by single pointings lasting 27.4 days. This relatively short observational coverage may lead to concerns regarding the precision of the ephemerides of the transiting objects found, since ephemeris with large uncertainties may make the re-observation of a planet (or candidate, it does not matter in this context) very difficult. This becomes the case once a predicted transit has timing errors larger than 2-3 hours, which makes the appearance of that transit in a given observing night rather uncertain. During the photometric follow-up of CoRoT planet candidates \citepads{2009A&A...506..343D}, this concern also surfaced, especially for candidates with shallow transits or for those found in `short-run' pointings of less than 50 days. It led to the implementation of a timing-precision estimator for all candidates \citepads{2012arXiv1206.1212D} and motivated follow-up observations of all detected planets, aimed at refining their ephemerides to a precision that maintains timing errors below 1 hour over at least one decade \citepads{2015EPJWC.10106020D}.
Table~\ref{tab:TESStiming} shows five groups of examples with the timing precision that can be expected from TESS. It also shows the corresponding precision of the ephemeris that may be obtained. The columns of that table are, from left to right, as follows: 
\begin{description}
\item[$R_\mathrm{pl}$, $R_*$:]{Sizes of the planet and the star.}
\item[$P_\mathrm{pl}$, $N_\mathrm{tr}$:]{Period of the planet and the number of transits that are observed by TESS, assuming its predominant temporal coverage of 27.4 days}
\item[$I_C$:]{Target magnitude in the $I_C$ band of TESS. This band is at the centre of the TESS passband}
\item[$\sigma_F$:]Expected 1-$\sigma$ photometric precision for a target with given $I_C$ magnitude over 1h integration time. This value was taken from the `total precision' for a given magnitude in Fig. 8 of \citetads{2015JATIS...1a4003R}.
\item[$\Delta F$:]Relative depth of a transit, assuming $\Delta F = (R_\mathrm{pl} / R_*)^2$ , e.g. without limb-darkening.
\item[$(S/N)_\mathrm{det}$:] S/N of the planet detection in the entire sequence of transits. This value is derived for a circular orbit of the planet around its central star, with the transit durations derived from Kepler's third law, assuming equatorial transits and a stellar mass given by the empirical mass-radius relation for main-sequence stars \citepads{1991Ap&SS.181..313D}: $M_*=(R_*/1.06)^{(1/0.945)}$.
\item[$T_\nabla$:]The duration of the combined ingress and egress time for an equatorial transit.
\item[$N_\nabla$:] Number of data points in $T_\nabla$, the sum of ingress and egress time. The TESS time-sampling of 2 minutes is used.
\item[$\sigma_{F_\nabla}$:] Expected 1-$\sigma$ photometric precision over the timescale of $T_\nabla$.
\item[$(S/N)_t$:] Timing signal-to-noise, as defined by Eq.~\ref{eq:SNt}, for a single transit.
\item[$\sigma_t$:] Timing precision of single transits observed by TESS, derived from Eq.~\ref{eq:TEEbasic} (or from Eq.~\ref{eq:TEEgen} with $\tau = 1h$).
\item[$\sigma_P/P$:]Relative period precision of the detected planet. $\sigma_P$ was calculated with the formula of \citetads{2015A&A...578A..17D} for the period-error in a continuous set of timing measurements, given by $\sigma_P = \sigma_t\  ({12}/{( N_\mathrm{tr}^3-N_\mathrm{tr})})^{1/2}$.
\item[$\sigma_{T,\mathrm{1yr}}$:] Precision of the planet's ephemeris 1 year after the first transit was detected by TESS. For this value,  Eq. 18 of \citetads{2015A&A...578A..17D}  was used, assuming $\sigma_t$ to be the ephemeris epoch (or zero-point) error. The dominant term of $\sigma_{T,\mathrm{1yr}}$ is, however, the period-error $\sigma_P$.
\end{description}
The first four cases (sample \emph{i}) correspond to planet detections at the limiting $I_C$ magnitude indicated by a revised version\footnote{During the preparation of this work, we noted a significant discrepancy between a detection S/N that was scaled from Sullivan et al.'s value of 7.3 and a detection S/N that was derived from first principles. After contacting Sullivan et al., they found an error in their Fig. 3 and provided a revised version with limiting magnitudes that are 1.5 - 3 mag brighter (Errata to Sullivan et al. 2015, in prep.). They indicated that the error that affected their Fig. 3 does not have consequences on any other results of their paper.} of Fig. 3 of \citetads{2015ApJ...809...77S}, who define  a detection  as ``achieving a signal-to-noise ratio greater than 7.3 from 6 hours of integration time during transits''. Since our assumptions led to somewhat longer total on-transit integration times, the $(S/N)_\mathrm{det}$ of the four examples given augments to values around 10. These four examples were selected to cover the range of stellar sizes that constitute the principal sample of TESS, and to cover the range of planet-sizes for which Sullivan et al. calculated limiting magnitudes. 

The next three samples \emph{ii - iv} are identical to sample \emph{i} except for the modification of a single parameter. In sample \emph{ii}, the target stars are 10 times (or 2.5mag) brighter; in sample \emph{iii} the radii of the planets  have been multiplied by 2.5 (the larger planets are now about Jupiter-sized), and in sample \emph{iv} the orbital period has been divided by 3, leading to three times as many transit events.  In sample \emph{v}, both the planet size and the target brightness have been increased relative to the cases at the detection limit.  

Regarding the interpretation of these calculations, first we note that in all cases, the derived timing precision are {lower} limits to the timing precision. This is due to the assumption of equatorial transits, where ingress/egress times are shortest, and is due to neglecting stellar limb-darkening, which also leads to ingress/egress times that may be shorter (but never longer) than real transit ingresses/egresses, in which the inner part often displays a notable rounding as well. Even with these shortest-possible ingress/egress times, we find that $N_\nabla$ is $\ga 3$ in all cases, meaning that the TESS two-minute sampling interval is sufficiently short to avoid any degradation of the timing precision.

For the cases at the faint detection limit (sample \emph{i}), their theoretical timing error $\sigma_t$ is 4 - 5 minutes, while the error of the ephemeris of such planets (or planet candidates) will be about 2 hours after one year. 
A further warning on the observability of these detections arises from their low timing signal-to-noise $(S/N)_t$. As our simulations in Sect. 3  show, true timing errors will exceed the theoretical errors for $(S/N)_t \la 10$.
Transits with timing errors of 2 hrs can usually still be recovered in single-night follow-up observations; but given above warning, a rapid follow-up within a few months is advisable, before multi-night or multi-site campaigns may become required in order to `recover' the transits of these detections at the limit of TESS.
  
For samples \emph{ii} and \emph{iii}, with a detection S/N of 20 - 60, we expect that one year after the observations by TESS, their ephemeris errors will be on the order of one hour. That is, a follow-up within about one year after their detection is advisable.
Sample \emph{iv} evaluates the gain from shorter orbital periods. The timing precision of individual transits ($\sigma_t$ in Table~\ref{tab:TESStiming}) improves only slightly over sample \emph{i}, due to the shorter ingress and egress times ($T_\nabla$). The larger number of observed transits,  however, leads  to an improvement in the ephemeris precision ($\sigma_P/P$ and $\sigma_{T,\mathrm{1yr}}$) by a factor of $\approx$ 2. For such planets, with a detection S/N of 10-15, follow-up observations should also be performed within one year.

Only for sample \emph{v}, with a detection S/N $>$ 200, will TESS derive ephemerides that are sufficiently precise to permit re-observation of their transits for several years. A refinement of these ephemerides by a dedicated follow-up program would however be desirable as well, in order to obtain ephemerides that are sufficiently precise for useful transit predictions over timescales of 10 years or more. 

\section{Conclusions}
In this  work we have introduced and validated the TEE, a formula that permits a quick estimation of the expected timing precision of individual transits or eclipses. This is expected to be useful for estimates of  timing precision for both ground- and space-based data. Together with the work by Deeg (2015) on period and ephemeris errors from time series covering multiple orbital periods, and assuming that there are no intrinsic period variations (e.g. transit timing variations, TTVs), we now have the tools at hand to estimate the precision of ephemerides for the majority of planet detections (those without TTVs) from rather basic assumptions, as is shown in the preceding example on TESS detections.

From the simulations shown in Sect. 3, we can conclude that the TEE accurately represents the timing precision of idealised trapezoidal eclipses, provided that the timing signal-to-noise is $(S/N)_t > 10$ and the number data points during ingress and egress, $n_\nabla$, is larger than 2. While Eq.~\ref{eq:TEEnorm} shows that no useful timing error can be measured for $(S/N)_{t} \la  1$, as $\sigma_{t}$ will exceed the length of an ingress or egress, in practice such low values of $(S/N)_t$ are not expected to occur as they would have originated from detections below usual detection thresholds. The TEE also predicts the timing error of naturally occurring transits, for which it is sufficient to obtain the  TEE's required parameters from trapezoidal fits to the transits. Moreover, the limitations on timing precision from low $n_\nabla$ seem to have a lesser effect on natural transits as they have, in general, no truly flat portions -- it is not the flat portion of a trapezoidal eclipse that contributes to the timing precision, but the slopes of the ingress/egress. 
We note that one of the assumptions in the TEE are negligible errors in its input parameters (e.g. the depth of the transit). Considering that the TEE does not estimate  a value but an error -- for which 1-2 digit precision is sufficient -- this assumption should not have consequences in practical applications. In any case, relevant uncertainties of the input parameters will occur only in the low S/N regime, where our simulations have already shown that the results from the TEE lead to underestimations of the real timing error. We emphasise here that the TEE is not intended to generate ‘final’ publication-quality timing errors if a full transit modelling is feasible.

The results from our simulations also have implications for the observational design of instruments acquiring photometric time series. As a general rule, eclipses or transits should be observed with cadences that are shorter than the expected ingress or egress duration (e.g. $n_\mathrm{in} > 1$ is desirable).  
Beyond this, and assuming that a white noise scaling is valid, the number of data points during ingress or egress has little effect  on the real timing precision or on the reliability of prediction made with the TEE.

As an example of this, the Kepler/K2 mission obtained most of its time series with its `long-cadence' (29.4 min) sampling. For eclipsing bodies that are small or have short periods, ingress/egress times shorter than 30 min are frequently encountered. The timing precision of these short-period objects would have been substantially better if they had been observed with cycle times a half or a quarter as long as the long cadence. For example, \citetads{2011ApJS..197....2F} %ford11
indicate an improvement in the timing error of the KOI-137 system (published in parallel as Kepler 18 by \citeads{2011ApJS..197....7C}%cochran
) of a factor of 2 for its short-cadence  ($\approx$ 1 min sampling) data relative to its long-cadence data. 

Regarding the TESS mission, our calculations on expected ephemeris precision have shown that for transiting planets found near the TESS detection limit, a photometric follow-up to acquire further transit times should be performed within one year after their initial detection, in order to maintain their future observability. Also, for the majority of its expected planet discoveries, further photometric follow-up will be required in order to establish ephemerides that guarantee their long-term observability.

This will  not be a worry with the PLATO space mission, foreseen to be launched by ESO in late 2025. This mission will observe large swaths of the sky with campaigns lasting 2-3 years \citepads{2014ExA....38..249R} while its nominal sampling rate of 25 seconds (with bright objects being sampled at a rate of 2.5 seconds) greatly exceeds the requirements for precision timing of any kind of normal planet-star system. This rapid sampling may have other benefits as well, such as a more reliable recovery of the shapes of eclipses from short-period orbiters or the precise timing of eclipses across compact objects, such as white dwarfs, where total eclipse durations of the order of minutes may be encountered.

\begin{acknowledgements}
      We thank the anonymous referee for comments that led to a significant improvement in the presentation of this work. HD thanks the Stellar Astrophysics Centre at Aarhus University for a visitor stay during which this project was conceived. He acknowledges also support by grant ESP2015-65712-C5-4-R from the Spanish Secretary of State for R\&D\&i (MINECO).
\end{acknowledgements}

%-------------------------------------------------------------------

\bibliographystyle{aa} % style aa.bst 
%\bibliographystyle{adsaa} %% adsaa.bst adds ADS bibcodes to references
%\bibliography{../../HJDmain,timingprecision5}
\bibliography{timingprecision7}

\end{document}